\documentclass[prl,aps,twocolumn,epsfig,showpacs,superscriptaddress,nofootinbib]{revtex4}
\usepackage{amsmath}
\usepackage{amssymb}
\usepackage[dvips]{graphicx}
\usepackage{dcolumn}
\usepackage{bm}
\usepackage[colorlinks=false,dvipdfm]{hyperref} 
\usepackage{setspace}
\usepackage{amsfonts}
\usepackage{rotating}
\usepackage{makeidx}
\usepackage{amsthm}

\newcommand{\tr}{\mbox{Tr}}

\newcommand{\Diag}{\mbox{Diag}}

\begin{document}

\title{Linear response of tripartite entanglement to infinitesimal noise \footnote{Annals of Physics (2014) In Press. \\http://dx.doi.org/10.1016/j.aop.2014.06.017}}

\author{Fu-Lin Zhang}
\email[]{flzhang@tju.edu.cn}
\affiliation{Physics Department, School of Science, Tianjin
University, Tianjin 300072, China}

\author{Jing-Ling Chen}
\email[]{chenjl@nankai.edu.cn}
\affiliation{Theoretical
Physics Division, Chern Institute of Mathematics, Nankai University,
Tianjin, 300071, China} \affiliation{Centre for Quantum
Technologies, National University of Singapore, 3 Science Drive 2,
Singapore 117543}

\date{\today}

\begin{abstract}
Recent experimental progress in prolonging the coherence time of a quantum system prompts us to explore the behavior of quantum
entanglement at the beginning of the decoherence process.
The response of the entanglement under an infinitesimal noise can serve as a signature of the robustness of entangled states.
A crucial problem of this topic in multipartite systems is to compute the degree of entanglement in a mixed state.
We find a family of global noise in three-qubit systems, which is composed of four W states.
Under its influence, the linear response of the tripartite entanglement of a symmetrical three-qubit pure state is studied.
 A lower bound of the linear response is found to depend completely on the initial tripartite and bipartite entanglement.
This result shows that the decay of tripartite entanglement is hastened by the bipartite one.
\end{abstract}



\maketitle



\emph{Introduction.--}
Dynamics of quantum entanglement in open systems has attracted wide attention in recent years,
 because of the key roles of entanglement in quantum information processing and unavoidable interaction of a quantum systems with its environment \cite{RevModPhys.81.865,RevModPhys.75.715,RevModPhys.76.1267}.
The coherence time of some quantum systems have been significantly extended in recent experimental
progress \cite{du2009preserving,PhysRevLett.105.053201},
 which makes it necessary to investigate the behavior of quantum
entanglement under a perturbation of decoherence.
In addition, the response of entanglement under an infinitesimal noise
 can be considered as a quantitative signature of the robustness of an
entangled state \cite{Zhang2013136}.
For the bipartite entanglement in a multipartite system measured by negativity \cite{NEG1}, which is an eigenvalue problem,
 the response can be treated by using the perturbation theory \cite{pertubation}.
 Consequently, the first author of the present article and his coworkers \cite{Zhang2013136} derive an analytical expression of the linear response of negativity (LRN) for an arbitrary three-qubit pure state in terms of its entanglement invariants \cite{sudbery}.
The roles of different initial entanglement in decay of bipartite entanglement are shown clearly in the analytical result.
  This naturally leads to an
interesting question: \emph{Can we find an analytical relation between the linear response of three-tangle \cite{coffman2000distributed} (LRT) in a three-qubit system in some noisy environment and its initial entanglement?}

Three-tangle is called
originally the residual entanglement quantifying
genuine multipartite correlations in a three-qubit pure state \cite{coffman2000distributed}.
It can be extended to mixed states via the convex roof \cite{PhysRevA.54.3824,PhysRevA.62.032307} as concurrence for the two-qubit case \cite{Wootters98} .
But the problem to compute the convex roof in an arbitrary three-qubit state has not been solved.
There are very limited examples of mixed states whose three-tangle have been obtained
analytically \cite{PhysRevA.77.032310,PhysRevA.80.010301,PhysRevA.79.024306,PhysRevLett.97.260502,eltschka2008three,PhysRevLett.108.230502}.
All of these examples can be considered as a GHZ or generalized GHZ state influenced by different types of global noises.
Lack of parameters in generalized GHZ states makes it impossible to analyze the influences of different entanglement on the decay of three-tangle.
In the present work, we consider the three-qubit system prepared in symmetrical pure states, which have three linear independent entanglement invariants \cite{sudbery}.
Two natural choices of noise environment are the partially depolarizing noise and global white noise, which are invariable under both local unitary transformations and permutations of qubits.
The former is studied in several works to explore the robustness of multipartite entanglement \cite{Zhang2013136,dur2004stability,aolita},
and the latter leads to the Werner states \cite{Werner1989,PhysRevLett.108.230502}.
However, the difficult of the optimal decomposition keeps us from moving forward in the two directions.
The question then is: \emph{Can we find a type of noise, the LRT corresponding to which not only can be derived but also has an explicit relationship with the initial entanglement?}


We begin from the simple case of symmetrical two-qubit pure states.
By assuming a form of optimal decomposition, we obtain the linear response of concurrence (LRC) under a family of \emph{W-type noise},
 which agrees with the result by using Wootters' formula \cite{Wootters98} and depends completely on the initial entanglement.
Subsequently we generalize these results to the three-qubit case.
The LRT, actually its lower bound because of the absence of a proof of the optimal decomposition, is found as a function of the initial bipartite negativity and three-tangle.


\emph{Two-qubit concurrence.--}
A two-qubit pure state can always be converted
into the following symmetric form with a suitable local unitary
operation
\begin{eqnarray}
|\Phi\rangle=\cos \theta |00\rangle + \sin\theta |11\rangle,
\end{eqnarray}
with $\theta \in[0,\pi/4]$. Its concurrence $\mathcal{C}(\Phi)=|\langle\Phi^* |\sigma_y \otimes \sigma_y  |\Phi\rangle|=\sin 2 \theta$, where $\sigma_{x,y,z}$ denote the Pauli matrices.
It orthogonal to the three linear independent two-qubit states as $|\Phi_0\rangle=\sin \theta |00\rangle - \cos \theta |11\rangle$, $|\Phi_1\rangle= |01 \rangle$, and  $|\Phi_2\rangle= |10 \rangle$.
Two properties of the latter two can be easily noticed: (i) $\mathcal{C}(\Phi_{1,2})=0$;
 (ii) for arbitrary $\varphi \in[0,2 \pi ]$ and $0<z \ll 1$, one has $\mathcal{C}(\Phi^{\prime}_{1,2}) <\mathcal{C}(\Phi)$, with $|\Phi^{\prime}_{1,2}\rangle =\sqrt{1-z}|\Phi\rangle+\sqrt{z} e^{i \varphi} |\Phi_{1,2}\rangle$.
The relationship between $|\Phi\rangle$ and $|\Phi_{1,2}\rangle$ is similar with the GHZ state and the W state for consideration three-tangle \cite{PhysRevA.77.032310}.
We refer to the white noise in the subspace of $\{|\Phi_{1}\rangle,|\Phi_{2}\rangle\}$ as \emph{W-type noise}, which is given by $\Lambda_W=(|\Phi_1\rangle \langle \Phi_1|+|\Phi_2\rangle \langle \Phi_2|)/2$.

Influenced by W-type noise, the two-qubit state becomes
\begin{eqnarray}
\rho_{\Lambda}=(1-q)|\Phi\rangle \langle \Phi| + q \Lambda_W,
\end{eqnarray}
where $q\in[0,1]$ denotes the strength of noise.
When $q \ll 1$, its concurrence in first-order approximation can be written as
\begin{eqnarray}
\mathcal{C}(\rho_{\Lambda})=\mathcal{C}(\Phi) - \eta_{\mathcal{C}}(\Phi) q ,
\end{eqnarray}
where $\eta_{\mathcal{C}}(\Phi)$ is the LRC of $|\Phi\rangle$ under W-type noise.
The concurrence for a
mixed state $\rho$
is defined as the average
pure-state concurrence minimized over all possible decompositions $\rho= \sum_{j}p_j |\phi _{j}\rangle \langle\phi _{j} |$,
\begin{eqnarray}
\mathcal{C}(\rho)=\min \sum_{j}p_j \mathcal{C}(\phi _{j}).
\end{eqnarray}

To calculate the concurrence for the two-qubit state $\rho_{\Lambda}$,
we assume the pure states in one of its optimal decomposition have the form
\begin{eqnarray}\label{pureform}
|\phi_j\rangle=\sqrt{1-q} |\Phi\rangle + \sqrt{q}( a_j |\Phi_1\rangle+ b_j |\Phi_2\rangle),
\end{eqnarray}
with $|a_j |^2 + |b_j |^2=1$. Its concurrence can be obtain directly as
$
\mathcal{C}(\phi_j)=|(1-q)\sin 2 \theta - 2 q a_j b_j |.
$
The last term corresponds to the coupling between the two state $|\Phi_1\rangle$ and $|\Phi_2\rangle$.
Choosing another basis $|\Phi_{\pm}\rangle=(|\Phi_1\rangle\pm|\Phi_2\rangle)/\sqrt{2}$, $ a_j |\Phi_1\rangle+ b_j |\Phi_2\rangle =A_j|\Phi_{+}\rangle+B_j |\Phi_{-}\rangle$ with $|A_j |^2 + |B_j |^2=1$ and
\begin{eqnarray}\label{pureC}
\mathcal{C}(\phi_j)=|(1-q)\sin 2 \theta -  q(A_j^2 + B_j^2) |.
\end{eqnarray}
All possible decompositions with the elements (\ref{pureform}) satisfy
$\sum_{j}p_j|A_j|^2=\sum_{j}p_j|B_j|^2=1/2$ and $\sum_{j}p_j A_j =\sum_{j}p_j B_j = \sum_{j}p_j A^*_j B_j=0$.
Among these sets of $\{p_j,A_j,B_j  \}$, for $p\ll1$, one can minimize $\sum_{j}p_j \mathcal{C}(\phi _{j})$ at $p_j=1/4$ $(j=1,2,3,4)$, $A_1=A_2=-A_3=-A_4=1/\sqrt{2}$, and $B_1=-B_2=B_3=-B_4=1/\sqrt{2}$.
The corresponding minimum average concurrence
\begin{eqnarray}
\min_{\{p_j,A_j,B_j  \}} \sum_{j}p_j \mathcal{C}(\phi _{j})= \sin 2 \theta -  q (1+\sin 2 \theta).
\end{eqnarray}
It actually is the exact concurrence
for the perturbed two-qubit state $\rho_{\Lambda}$, which can be checked by using  Wootters' formula \cite{Wootters98}.
The LRC can be written as
\begin{eqnarray}\label{LRC}
 \eta_{\mathcal{C}}(\Phi) = \mathcal{C}(\Phi)+1,
\end{eqnarray}
which depends completely on the initial entanglement.
One can notice from  (\ref{pureC}) that the concurrence $\mathcal{C}(\Phi)$ in $\eta(\Phi)$
 comes from the decrease of the pure state $|\Phi\rangle$, and the constant $1$ from the coupling between the two state $|\Phi_1\rangle$ and $|\Phi_2\rangle$ in W-type noise.
In addition, the constant $1$ can be also considered as a trivial entanglement invariant of two-qubit pure state $|\Phi\rangle$.
This viewpoint will be supported by the results for three-tangle in the next section.

\emph{Three-tangle.--}
Now we turn to consider the symmetrical three-qubit pure states.
These states have
three independent parameters and, through an appropriate local unitary
transformations, can be brought into the compact form \cite{PhysRevA.82.032301}
\begin{eqnarray}\label{pure3}
|\Psi\rangle=\cos \alpha |\bar{W}\rangle+\sin\alpha (\cos\beta |000\rangle +   \sin\beta e^{i \gamma}|111\rangle),
\end{eqnarray}
where $\alpha,\beta \in [0,\pi/2]$, $\gamma \in[-\pi/2,\pi/2]$,
 and $|\bar{W}\rangle=\sigma_x^{\otimes 3} | W \rangle$ is a flipped version of the original W state $|W \rangle=(|001\rangle+|010\rangle+|100\rangle)/\sqrt{3}$.
The tripartite entanglement of a three-qubit pure state $|\psi\rangle$ is defined as the three-tangle \cite{coffman2000distributed}
\begin{eqnarray}\label{tangle}
\tau (\psi)= \biggr|\sum_{j=0,x,z}\langle \psi^*|\sigma_j\otimes \sigma_y\otimes\sigma_y |\psi\rangle\langle \psi^*|\sigma_j\otimes \sigma_y\otimes\sigma_y |\psi\rangle \biggr|,  \ \ \ 
\end{eqnarray}
with $\sigma_0 = i \openone_2$.
For the symmetrical state $|\Psi\rangle$ in (\ref{pure3}), it is given by
\begin{eqnarray}
\tau (\Psi)= |\mathcal{T} (\alpha,\beta,\gamma)|,
\end{eqnarray}
where we define the complex number $\mathcal{T} (\alpha,\beta,\gamma)=   (16\sqrt{3}/9)\cos^3 \alpha \sin \alpha \cos \beta + 4 e^{2 i \gamma}\sin^4 \alpha \cos^2 \beta \sin^2 \beta  $.

There exist four linear independent states $|\Psi_k\rangle$ $(k=1,2,3,4)$  orthogonal to $|\Psi\rangle$  satisfying the similar properties as the case of two-qubit concurrence: (i) $\tau(\Psi_k)=0$; (ii)  $\tau(\Psi^{\prime}_k) < \tau (\Psi) $ for $|\Psi^{\prime}_{k}\rangle =\sqrt{1-z}|\Psi\rangle+\sqrt{z} e^{i \varphi} |\Psi_{k}\rangle$ with  $\varphi \in[0,2 \pi ]$ and $0<z \ll 1$.
They are $|\Psi_1\rangle=  e^{i 2\pi/3 \sigma_z} \otimes e^{-i 2\pi/3 \sigma_z } \otimes \openone_2  |W\rangle$, $|\Psi_2\rangle= e^{-i 2\pi/3 \sigma_z} \otimes e^{i 2\pi/3 \sigma_z } \otimes \openone_2   |W\rangle$,
$|\Psi_3\rangle= \openone_2 \otimes  e^{i 2\pi/3 \sigma_z} \otimes e^{-i 2\pi/3 \sigma_z }  |\bar{W}\rangle$, and $|\Psi_4\rangle= \openone_2 \otimes  e^{-i 2\pi/3 \sigma_z} \otimes e^{i 2\pi/3 \sigma_z }  |\bar{W}\rangle$.

Applying the three-qubit W-type noise $\Pi_W=\sum_k | \Psi_k \rangle \langle \Psi_k|/4 $ with a infinitesimal probability $q$,
we obtain the three-qubit state initial from $|\Psi\rangle$ as
\begin{eqnarray}\label{mixed3}
\rho_{\Pi}=(1-q)|\Psi\rangle \langle \Psi| + q \Pi_W.
\end{eqnarray}
In order to derive its three-tangle defined by
the convex roof of $\tau$, we extend the assumed optimal pure states (\ref{pureform}) into the three-qubit case as
\begin{eqnarray}\label{pureform3}
|\psi_j\rangle=\sqrt{1-q} |\Psi\rangle + \sqrt{q} \sum_k c^j_k |\Psi_k\rangle,
\end{eqnarray}
with $\sum_k |c^j_k |^2  =1$.
Its three-tangle in the first-order approximation is given by
\begin{eqnarray}\label{tangleansatz}
\tau (\psi_j)= |(1-2q)\mathcal{T} (\alpha,\beta,\gamma) - q \langle \chi_j^*| \Omega  |\chi_j\rangle|,
\end{eqnarray}
where $|\chi_j\rangle = (c^j_1,c^j_2,c^j_3,c^j_4)^{\rm T}$ and $\Omega$ is a $4\times4$ symmetric matrix depending on the parameters $\alpha,\beta,\gamma$.
Choosing a suitable basis $\{ |\tilde{\Psi}_k\rangle \}$, $\sum_k c^j_k |\Psi_k\rangle =  \sum_k C^j_k |\tilde{\Psi}_k\rangle$, the coupling term in (\ref{tangleansatz}) can be diagonalized
as $\langle \chi_j^*| \Omega  |\chi_j\rangle = \sum_k \omega_k (C^j_k)^2 $ \cite{DiagMatrix}.
 For fixed absolute values $|C^j_k|$, the minimum of $\tau (\psi_j)$ is $\tau_{\min}(\psi_j) = (1-2q)\tau(\Psi)- q\sum_k |\omega_k| |C^j_k|^2$.
Here, their convex combination also achieves its minimal value
\begin{eqnarray}\label{tangleMin}
\min_{\{p_j, C^j_k \}}\sum_j p_j \tau (\psi_j)&=& \sum_{j}p_j\tau_{\min}(\psi_j) \nonumber\\
&=&(1-2q)\tau(\Psi)- \frac{q}{4}\sum_k |\omega_k|, \ \ \
\end{eqnarray}
 where we utilized the constraint $\sum_{j}p_j|C^j_k|^2=1/4$.
In addition, a possible decomposition $\{p_j, C^j_k \}$ should satisfies $\sum_j p_j C^j_k=\sum_j p_j C^j_k (C^j_{l \neq k})^*=0$ and $\sum_k |C^j_k|^2 =1$.
A optimal decomposition $\{p_j, C^j_k \}$ reaching the minimum in (\ref{tangleMin})
can be constructed as $C^j_k=\frac{1}{2}e^{i(\xi + \zeta_k+\delta)}$, with $\xi=\frac{1}{2}\arg \mathcal{T} (\alpha,\beta,\gamma)$,  $\zeta_k=\frac{1}{2}\arg \omega_k $, and $\delta=0$ or $\pi$.
 Each of the $2^4$ sets of $\{C^j_k \}$ corresponds to a probability $p_j=1/2^4$.
In this way, we minimize the average three-tangle over all decompositions with the elements in (\ref{pureform3}).
That is, the minimum in (\ref{tangleMin}) is a upper bound of the three-tangle for the perturbed state $\rho_{\Pi}$.
Consequently, a lower bound of the LRT is obtained as $ \eta_{\tau} (\Psi)= 2 \tau(\Psi) + \sum_k |\omega_k|/4 $.
In the following part, we treat it as the exact LRT, and will discuss its correctness in the end of the article.

Our goal is now to derive the values of $|\omega_k|$ in (\ref{tangleMin}).
The basis transformation to $\{ |\tilde{\Psi}_k\rangle \}$ can be expressed as a $4\times4$ unitary operation $|\mu_j\rangle = (C^j_1,C^j_2,C^j_3,C^j_4)^{\rm T}= U |\chi_j\rangle$.
The relation $ \langle \mu_j^* | D |\mu_j\rangle = \langle \chi_j^*| \Omega  |\chi_j\rangle$ leads to $U^{\rm T} D U =  \Omega$, where the diagonal matrix
$D= \Diag (\omega_1,\omega_2,\omega_3,\omega_4)$.
One can easily notice $\Omega^*  \Omega =U^{\dag} D^*   D U=U^{\dag} \Diag (|\omega_1|^2,|\omega_2|^2,|\omega_3|^2,|\omega_4|^2) U$.
Therefore, the values of $|\omega_k|$ are the square roots
of the eigenvalues of the  Hermitian matrix $R=\Omega^*  \Omega$, which has the form as
\begin{eqnarray}\label{Rmatix}
R=\left[
\begin{array}{cccc}
X & 0 & 0 & Z_+ \\
 0 & X  & Z_-  & 0 \\
 0 & Z^*_-  & Y  & 0 \\
Z^*_+ & 0 & 0 & Y
\end{array}
\right]. \ \
\end{eqnarray}
Here, the elements $X= \frac{64}{9} \cos^4 \alpha  +4\sin^4\alpha \sin^2 2\beta$, $Y=\frac{64}{3} \cos^2 \alpha \sin^2 \alpha \cos^2 \beta + 16\sin^4 \alpha \cos^2 \beta \sin^2 \beta $, and $Z_{\pm}=\frac{32}{3}e^{i(-\gamma \pm \pi/3)}\cos^2 \alpha \sin^2 \alpha \sin^2 \beta - \frac{32}{\sqrt{3}}e^{i(\gamma \pm \pi/3)} \cos\alpha \sin^3 \alpha \cos^2 \beta \sin \beta$.
The traces and determinants of the two submatrice in the partitioned matrix $R$ are given by
$
X+Y=8 \mathcal{N}^2(\Psi)$ and $
XY-|Z_+|^2 = XY-|Z_-|^2 = 16 \tau^2 (\Psi),
$
where $\mathcal{N}(\Psi) = \sqrt{4 \det  \rho_1 }$ is the bipartite negativity between one qubit with the other two, and $\rho_1=\tr_{2,3} |\Psi\rangle\langle \Psi| $ is the one-qubit reduced state of $|\Psi\rangle$.

Based on the matrix $R$ in (\ref{Rmatix}), it is straightforward to derive the LRT
\begin{eqnarray}\label{LRT}
\eta_{\tau} (\Psi) = 2 \tau(\Psi) + \sqrt{ \mathcal{N}^2(\Psi) -\sqrt{ \mathcal{N}^4(\Psi)-\tau^2 (\Psi) }}&& \nonumber \\
+ \sqrt{ \mathcal{N}^2(\Psi) + \sqrt{ \mathcal{N}^4(\Psi)-\tau^2 (\Psi) }}&.&
\end{eqnarray}
Similar as the two-qubit concurrence, the first term is from the decrease of the pure states $|\Psi\rangle$.
The factor $2$ is caused by the fact that the three-tangle in (\ref{tangle})
is a
homogeneous function of degree $4$ in the coefficients of the pure states.
The last two terms from the addition of the noise become a function of the initial bipartite and tripartite entanglement
 advisors instead of a constant.
An interesting fact can be noticed is the absence of the Kempe invariant \cite{sudbery} in the LRT.

\begin{figure}
\centering
\includegraphics[width=7cm]{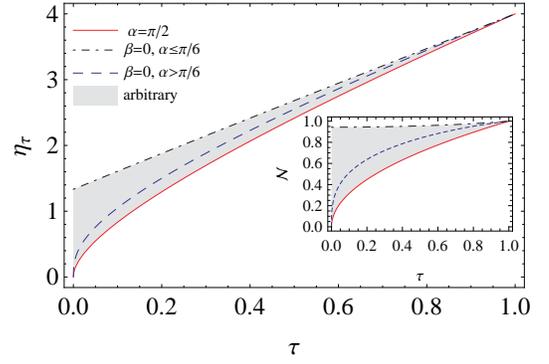} \\
\caption{The relation between the LRT
and the initial three-tangle of a three-qubit system in a generalized GHZ state (solid line),
a W-like state (dot-dashed line for $\alpha\leq\pi/6$ and dashed for $\alpha>\pi/6$), with the gray region for
arbitrary symmetrical pure state. The inset shows the their positions in the plane of bipartite vs. tripartite entanglement.
} \label{fig1}
\end{figure}

The analytical relation between the LRT and initial entanglement make it easy to identify the most robust and the fragile three-partite entanglement among
the symmetrical three-qubit pure states.
For a fixed three-tangle, the symmetrical states with minimal and maximal negativity, as shown in the inset in Fig. \ref{fig1}, are
the generalized GHZ state
\begin{eqnarray}
|G\rangle=|\Psi(\alpha=\frac{\pi}{2},\beta,\gamma)\rangle=\cos \beta |000\rangle + \sin \beta e^{i \gamma} |111\rangle, \ \ \
\end{eqnarray}
and the W-like state
\begin{eqnarray}
|J\rangle=|\Psi(\alpha,\beta=0,\gamma)\rangle=\cos \alpha |\bar{W}\rangle + \sin \alpha  |000\rangle,\ \ \
\end{eqnarray}
in the region of $\alpha\in[0,\pi/6]$.
They are also shown in Fig. \ref{fig1} as the state with the minimum and the maximum of LRT.
That is, the bipartite entanglement in a three-qubit state speeds up the decay of the tripartite entanglement.
Combining this point with the results in \cite{Zhang2013136}, we can say that the two kind entanglement, tripartite and bipartite,
are of a tendency to do harm to each other.
It is interesting to notice that when three-tangle approaches zero, LRT for most of the symmetrical states approaches a finite value, which is $\eta_{\tau \rightarrow 0}= \sqrt{2}\mathcal{N}$ and
with a maximum $4/3$ achieved by the W-like state.
This indicates a \emph{sudden death} of the slight three-tangle occurs in these states under a perturbation of the W-type noise.
These sudden death are obviously correlated with the bipartite negativity.
Only the fully separable states have a zero LRT.

\emph{Summary and discussion.-- }
We find a family of global noise, called W-type in the present paper
whose one-order perturbation on the concurrence of a symmetrical two-qubit pure state depends completely on initial entanglement.
To derive the perturbed concurrence, we present a ansatz for the optimal decomposition which conforms to the result by using Wotters' formula.
Subsequently we extend the noise and ansatz optimal decomposition to the three-qubit system prepared in a symmetrical pure state.
A lower bound of the LRT is found as an analytical function of the initial bipartite and tripartite entanglement.
It reveals that the bipartite entanglement between one qubit and the other two plays a role in speeding up the decay of tripartite entanglement.

\begin{figure}
\centering
\includegraphics[width=7cm]{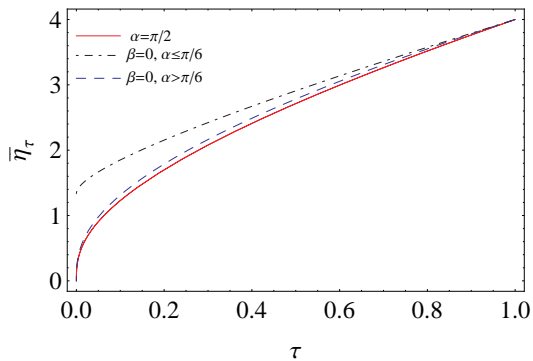} \\
\caption{
Average decay rates of three-tangle for generalized GHZ states (solid line) and W-like states (dot-dashed line for $\alpha\leq\pi/6$ and dashed for $\alpha>\pi/6$) under W-type noise.
} \label{fig2}
\end{figure}

Finally, we briefly discuss the validity of the lower bound as signature to characterize the roles of initial entanglement in decay of three-tangle.
The minimal average three-tangle among our ansatz decompositions is actually the minimized pure-state three-tangle $\tilde{\tau}$ in the method of \emph{convex
characteristic curves} \cite{PhysRevA.77.032310}.
We successfully find the decomposition achieving $\tilde{\tau}$ but fail to construct its function convex hull $\tau^*$.
One can notice from the known results \cite{PhysRevA.77.032310,PhysRevA.79.024306,PhysRevLett.97.260502,eltschka2008three}
 that the minimal value $\tilde{\tau}$ are all quite close to the exact three-tangle.
In addition, we derive the critical noise parameter $q_c$ (see Appendix) washing out initial tripartite entanglement of the states $|G\rangle$ and $|J\rangle$
 under the W-type noise by using the rescaling method \cite{viehmann2012rescaling}, and plot their average decay rates $\bar{\eta}_{\tau}(|\Psi\rangle)=\tau(|\Psi\rangle)/q_c$ in Fig. \ref{fig2}.
The relative positions of the states $|G\rangle$ and $|J\rangle$ are similar as the results in Fig. \ref{fig1}.
And, when the initial tripartite entanglement approach zero or one, the average decay rates tend to the values of LRT in (\ref{LRT}).
Based on these results, a same conclusion can be drawn that the bipartite entanglement makes three-tangle more fragile.

\begin{acknowledgments}
We thank Jens Siewert for his valuable comments.
F.L.Z. is supported by NSF of China (Grant No. 11105097). J.L.C. is
supported by National Basic Research Program (973 Program) of China
under Grant No. 2012CB921900, NSF of China (Grant Nos. 10975075 and
11175089) and also partly supported by National Research Foundation
and Ministry of Education, Singapore.
\end{acknowledgments}

\bibliography{TangleWNoise}

\appendix

\emph{Appendix.--}
In this part, we give the critical noise parameter where occurs the sudden death of tripartite entanglement in a three-qubit system prepared in states $|G\rangle$ or $|J\rangle$ under the W-type noise. Under local operation $A=[x |0\rangle \langle 0 | + (1/x) |1\rangle \langle 1 |] ^{\otimes 3}$ and renormalization, the noised state (\ref{mixed3}) can be transformed into
\begin{eqnarray}\label{mixed31}
\tilde{\rho}_{\Pi}=\frac{A\rho_{\Pi}A^{\dag}}{\tr( A\rho_{\Pi}A^{\dag})}=(1-\tilde{q})|\tilde{\Psi}\rangle \langle\tilde{ \Psi}| + \tilde{q} \Pi_W (p),
\end{eqnarray}
where $|\tilde{\Psi}\rangle= A |\Psi\rangle/\sqrt{\langle \Psi | A^{\dag} A |\Psi\rangle}$ and $\Pi_W(p)=[p(| \Psi_1 \rangle \langle \Psi_1|+| \Psi_2 \rangle \langle \Psi_2|)+(1-p)(| \Psi_3 \rangle \langle \Psi_3|+| \Psi_4 \rangle \langle \Psi_4|)]/2 $.
The parameters in $\tilde{\rho}_{\Pi}$ can be determined by
\begin{eqnarray}\label{pqx}
&&\tilde{q}=\frac{q}{2 \tr( A\rho_{\Pi}A^{\dag})} (|x|^2 + \frac{1}{|x|^2}), \nonumber \\
&&p=|x|^4 (1-p).
\end{eqnarray}
The three-tangle is a polynomial $SL(2,\mathbb{C})^{\otimes 3} $ invariant \cite{coffman2000distributed} with homogeneous degree 4.
A sufficient condition for $\tau(\rho_{\Pi})=0$ is $\tau(\tilde{\rho}_{\Pi})=0$ \cite{viehmann2012rescaling}.
Therefore, one can substitute the critical noise parameter $\tilde{q}_c$ of state $\tilde{\rho}_{\Pi}$ into the relations (\ref{pqx}) and
obtain the one of  $\rho_{\Pi}$.

Now, we choose $x= (\tan \beta e^{i \gamma})^{1/6}$ and $(\cot \alpha/\sqrt{3})^{1/4}$ for the initial states $|G\rangle$ and $|J\rangle$ respectively,
which are transformed into $A |G\rangle =\tau^{\frac{1}{4}}(G) |\tilde{G}\rangle$ and $A |J\rangle =\tau^{\frac{1}{4}}(J)|\tilde{J}\rangle$.
Here, the two pure states
\begin{eqnarray}
&& |\tilde{G}\rangle =  \frac{1}{\sqrt{2}}(|000\rangle +|111\rangle), \nonumber \\
&& |\tilde{J}\rangle = \frac{1}{2}(|000\rangle +|110\rangle+|101\rangle+|101\rangle),\ \
\end{eqnarray}
are equivalent under local unitary transformation and have the maximal tripartite entanglement.
According the method of convex
characteristic curves \cite{PhysRevA.77.032310}, we represent a pure state as
\begin{eqnarray}
|\psi\rangle=\sqrt{1-\tilde{q}} |\tilde{\Psi}\rangle  + \sqrt{\tilde{q}}\bigr( \sqrt{p}d_1 | \Psi_1 \rangle  + \sqrt{p} d_2 | \Psi_2 \rangle \ \ \ &&  \nonumber \\
+\sqrt{1-p}d_3 | \Psi_3 \rangle  + \sqrt{1-p} d_4 | \Psi_4 \rangle \bigr), &&
\end{eqnarray}
with real parameters $\tilde{q},p \in[0,1]$ and complex ones $|d_1|^2 + |d_2|^2=1$ and $|d_3|^2 + |d_4|^2=1$.

\begin{figure}
\centering
\includegraphics[width=6cm]{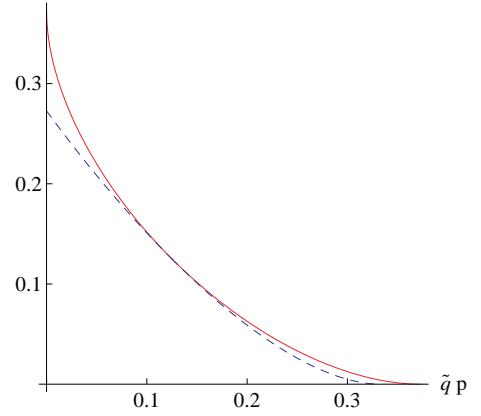} \\
\caption{
Critical noise parameters of $\tilde{\rho}_{\Pi}$ for $|\tilde{\Psi}\rangle=|\tilde{G}\rangle$ and $|\tilde{\Psi}\rangle=|\tilde{J}\rangle$
 are shown by solid and dashed curves respectively.
} \label{fig3}
\end{figure}

When $|\tilde{\Psi}\rangle=|\tilde{G}\rangle$, for fixed $\tilde{q}$ and $p$, the minimal three-tangle of state $|\psi\rangle$ is
$\tilde{\tau}_{G}= \max \{0,9 (1-\tilde{q})^2-12 (1-p) p \tilde{q}^2-8 \sqrt{6}\sqrt{1- \tilde{q}} \tilde{q}^{3/2}(1-p )^{3/2}-2 \sqrt{p}  [18 \sqrt{1-p} (1-\tilde{q}) \tilde{q}+4 p \sqrt{6}\sqrt{1- \tilde{q}} \tilde{q}^{3/2} ]\}/9$.
The critical point of the piecewise function is exactly the value of $\tilde{q}_c$ we are looking for.
This can be obtained from the two facts: (i) There exist six pure states $|\psi\rangle$ achieving the value of $\tilde{\tau}_{G}$ in the region of $\tilde{q}\in[0,\tilde{q}_c]$, which can be
written in two types as
\begin{eqnarray}
|\psi\rangle=\sqrt{1-\tilde{q}} |\tilde{G}\rangle  + \sqrt{\tilde{q}}\bigr( \sqrt{p} e^{i \theta_1} | \Psi_1 \rangle  +\sqrt{1-p} e^{i \theta_3}| \Psi_3 \rangle  \bigr),&& \nonumber \\
|\psi\rangle=\sqrt{1-\tilde{q}} |\tilde{G}\rangle  + \sqrt{\tilde{q}}\bigr( \sqrt{p}  e^{i \theta_2} | \Psi_2 \rangle +  \sqrt{1-p}  e^{i \theta_4} | \Psi_4 \rangle \bigr),&&\ \ \
\end{eqnarray}
where $\theta_1,\theta_2 \in \{\pi/3,\pi,-\pi/3 \}$, $\theta_3=2\pi/3 -\theta_1$ and $\theta_4=-2\pi/3 -\theta_2$.
These states with equal probabilities compose the state $\tilde{\rho}_{\Pi}$.
(ii) As shown in Fig. \ref{fig3}, the critical curve in the space of $(\tilde{q}p,\tilde{q}(1-p))$ is convex.

For the case of $|\tilde{\Psi}\rangle=|\tilde{J}\rangle$, the minimal three-tangle is
$\tilde{\tau}_{J}= \max \{0,9-36 \tilde{q}+3 (9-4 p+4 p^2) \tilde{q}^2-4 \sqrt{6} \sqrt{(1-p) (1-\tilde{q}) \tilde{q}^3}-8 \sqrt{6} p \sqrt{(1-p) (1-\tilde{q}) \tilde{q}^3}\}/9$.
Six pure states reaches the second part, which are
\begin{eqnarray}
|\psi_j\rangle=\sqrt{1-\tilde{q}} |\tilde{J}\rangle  + \sqrt{\frac{\tilde{q}}{2}}\biggr[ \sqrt{p}e^{i \delta^{(j)}_1} | \Psi_1 \rangle  + \sqrt{p} e^{i \delta^{(j)}_2} | \Psi_2 \rangle  \ \ \ \ &&  \nonumber \\
+\sqrt{1-p}e^{i \delta^{(j)}_3} | \Psi_3 \rangle  + \sqrt{1-p} e^{i \delta^{(j)}_4} | \Psi_4 \rangle \biggr], \ \ \ &&
\end{eqnarray}
with $j=0,...,5$, $\delta^{(j)}_1=\pi- \delta^{(j)}_2= j \pi/3$, and $\delta^{(j)}_3=-\delta^{(j)}_4=   2 \delta^{(j)}_1 -\pi/3$.
They satisfy $\sum_{j=0}^5|\psi_j\rangle \langle\psi_j |/6= (1-\tilde{q})|\tilde{J}\rangle \langle\tilde{J}| + \tilde{q} \Pi_W (p)$.
And as shown in Fig. \ref{fig3}, it is easy to find the zero points of $\tau(\psi_j)$ form a convex curve in the space of $(\tilde{q}p,\tilde{q}(1-p))$.
Therefore, the critical value $\tilde{q}_c$ of $\tilde{\tau}_{J}$ is precisely the noise parameter washing out initial three-tangle.

\end{document}